\documentclass[aps,prl,twocolumn,groupedaddress,showpacs]{revtex4}

\usepackage{latexsym} \usepackage{amssymb} \usepackage{amsmath}
\usepackage[dvips]{graphicx}
\usepackage[dvips]{color}

\begin{document}


\title{Rf-induced persistent long-range ordered structures in
  two-species ion Coulomb crystals in a linear Paul trap.}

\author{A.~Mortensen} \author{E.~Nielsen} \author{T.~Matthey}
\author{M.~Drewsen}\email[E-mail: ]{drewsen@phys.au.dk}

\affiliation{QUANTOP - Danish National Research Foundation Center for
  Quantum Optics, \\
  Department of Physics and Astronomy, University of Aarhus, DK-8000
  Aarhus C, Denmark}


\date{\today}

\begin{abstract}

We report on the observations of three-dimensional long-range ordered
structures in the central $^{40}$Ca$^+$ ion component of
$^{40}$Ca$^+$--$^{44}$Ca$^+$ two-species ion Coulomb crystals in a
linear Paul trap. In contrast to long-range ordering previously
observed in single species crystals, the structures observed are
strikingly persistent and always of one specific type in one
particular orientation. Molecular dynamics simulations strongly
indicate that these characteristics are a hitherto unpredicted
consequence of the co-axial cylindrical symmetry of the central ion
component of the Coulomb crystal and the radio frequency quadrupole trapping field.

\end{abstract}
\pacs{32.80.Pj, 52.27.Jt, 52.27.Gr, 36.40.Ei}
\maketitle

A solid state of an one-component plasma (OCP), often referred to as a
Wigner crystal or a Coulomb crystal, appears whenever the coupling
parameter $\Gamma=E_\mathrm{Coul}/E_\mathrm{kin}$, where
$E_\mathrm{Coul}$ is the nearest neighboring Coulomb potential energy
and $E_\mathrm{kin}$ is the averaged kinetic energy of the particles,
exceeds $\sim$200~\cite{Pollock:1973,Schiffer:2002,Dubin:1989}. In
recent decades, such crystals have been investigated experimentally in
a large variety of physical
systems~\cite{Birkl:1992,Grimes:1979,Andrei:1988,Thomas:1994}. In
particular, laser-cooled and trapped atomic ions have proven to be an
excellent system for experimental studies of Coulomb crystals under
various confinement conditions.  Single-species, one-, two- and
three-dimensional ion Coulomb crystals have been investigated in
Penning and radio frequency (rf) traps (also named Paul traps), with
structural findings in good agreement with theoretical
predictions~\cite{Birkl:1992,Drewsen:1998,Hornekaer:2001,Schatz:2001,Mitchell:1998,Itano:1998,Mortensen:2006}.
Two-species crystals have been studied in much less detail, and
experimentally mainly in linear Paul
traps~\cite{Hornekaer:2001,Blythe:2005}. For singly charged ions in
such traps, the lighter species will in general segregate into a
cylindrically shaped structure surrounded by the heavier ion species.
In the case of $^{40}$Ca$^+$ and $^{24}$Mg$^+$ ions, the structures of
the lighter species~\cite{Hornekaer:2001} were found for the most part
to be identical to the cylindrical structures of infinitely long
one-species systems confined in two-dimensions by a
rotational-symmetric harmonic potential~\cite{Hasse:1990}. For two ion
species with identical charge-to-mass ratios, mixing of the species is
predicted~\cite{Matthey:2003}.  At present, Coulomb crystals find
applications in such diverse fields as quantum
computing~\cite{Riebe:2004,Barrett:2004} and cold molecule ion
research~\cite{Molhave:2000,Blythe:2005,Bertelsen:2006}.

 \begin{figure}[htbp]
   \centering
   \includegraphics[width=8cm,clip=true]{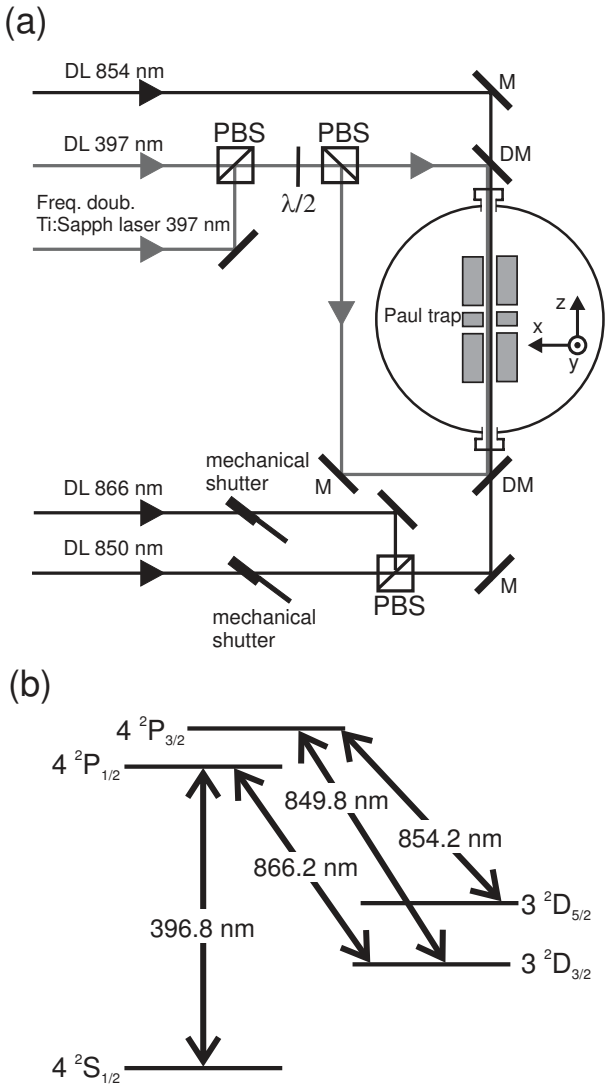}
   \caption{(a) Schematics of the laser and linear Paul trap
     setup. (D)M denotes a (dichroic) mirror, PBS denotes polarization
     beamsplitters, and DL denotes diode laser. (b) Optical
     transitions used for Doppler laser cooling of the Ca$^+$ ions.}
   \label{fig:figure1}
 \end{figure}
 In this Letter, we present observations of persistent
 long-range ordered structures in the $^{40}$Ca$^+$ component of
 bicrystals consisting of $^{40}$Ca$^+$ and $^{44}$Ca$^+$ ions. 
   The observed structures deviate from those expected  to exist in a
   fully rotational symmetric harmonic
   potential~\cite{Hornekaer:2001,Hasse:1990}, but are in close
   agreement with results from molecular dynamics (MD) simulations
   including the quadrupole nature of the trapping fields.
 
 The Ca$^+$ ions are confined in a linear Paul trap which is situated
 in a vacuum chamber at a pressure of $10^{-10}$ Torr, and are laser
 cooled as depicted in Fig.~1(a). The Paul trap used in these
 experiments has been described in detail
 elsewhere~\cite{Drewsen:2003}. In short, the linear Paul trap
 consists of four electrode rods placed in a quadrupole configuration.
 The electrode diameter is 8.0 mm and the minimum distance of the central
 trap axis is $r_0=3.5$ mm.  Time varying voltages
 $\frac{1}{2}U_\mathrm{rf}\cos (\Omega_\mathrm{rf} t )$ and
 $\frac{1}{2}U_\mathrm{rf} \cos(\Omega_\mathrm{rf} t +\pi)$ are
 applied to the two sets of diagonally opposite electrode rods,
 respectively, to obtain confinement in the radial plane ($xy$-plane
 in Fig.~1(a)). Axial confinement along the $z$-axis is accomplished
 by sectioning each of the electrode rods into three pieces and then
 applying a static voltage $U_\mathrm{end}$ to the end-electrodes. The
 length of the center-electrode is 5.4 mm, while the outer pieces are
 20 mm. In the present experiments $\Omega_\mathrm{rf} = 2\pi \times
 3.88$ MHz, $U_\mathrm{rf} \sim 540$ V and $U_\mathrm{end} \sim$
 10--50 V were used. The resulting confinement for an ion species of
 type $i$ with mass $M_i$ and charge $Q_i$ is  often approximated by a
 harmonic pseudo potential $\Phi_\mathrm{ps}(r,z) =
 \frac{1}{2}M_i(\omega_\mathrm{r}^2 r^2 +\omega_\mathrm{z}^2 z^2)$,
 where $\omega_\mathrm{r}$ and $\omega _\mathrm{z}$ are the radial and
 axial trap frequencies, respectively.  The axial trap frequency is
 given by $\omega_z^2 = 2\kappa Q_i U_\mathrm{end} /M_i$, where
 $\kappa = 3.97\times 10^4$ m$^{-2}$ is a constant related to trap
 geometry, and the radial trap frequency is given by $\omega_r^2 =
 \omega_{\mathrm{rf}}^2 - \frac{1}{2} \omega_z^2$, where $
 \omega_{\mathrm{rf}}^2 = Q_i^2U_\mathrm{rf}^2/ 2M_i r_0^4
 \Omega_\mathrm{rf}^2$ is the contribution from the time varying
   quadrupole
  fields.  The dependence on the charge and mass of the ion
 species makes the lighter isotope ($^{40}$Ca$^+$) more tightly bound
 towards the trap axis than the heavier ($^{44}$Ca$^+$) and leads
 consequently to a total radial separation of the two ion species when
 sufficiently cooled~\cite{Hornekaer:2001}. The zero-temperature ion
 density in the pseudopotential is given by $n_\mathrm{theo} =
 \epsilon_0 U_\mathrm{rf}^2/M_i r_0^4\Omega_\mathrm{rf}^2$, where
 $\epsilon_0 $ is the vacuum permittivity~\cite{Hornekaer:2001}. Due
 to the spatial separation of the ions in two-species ion Coulomb
 crystals, this expression is also applicable to the individual
 components of such crystals.

 The $^{40}$Ca$^+$ and $^{44}$Ca$^+$ ions used in the experiments are
 produced isotope selectively by resonant two-photon photo-ionization
 of atoms in an effusive beam of naturally abundant calcium
 ~\cite{Kjaergaard:2000,Mortensen:2004}.  In Fig.~1(b), the
 transitions in Ca$^+$ used for Doppler laser cooling of the trapped
 ions are shown. The main cooling transition is the dipole allowed
 $4S_{1/2} \to 4P_{1/2}$ transition at 397 nm. To avoid optical
 pumping into the metastable $3D_{3/2}$ state, repumping is done
 either by using a single repump laser at 866 nm via the $4P_{1/2}$ state
 ($^{44}$Ca$^+$) or by using two repump lasers at 850 nm and 854 nm via
 the $4P_{3/2}$ state ($^{40}$Ca$^+$). Due to the isotope shifts of
 the cooling transitions~\cite{Maartensson:1992,Alt:1997}, each isotope
 ion requires its own laser cooling frequencies. 
 For both isotopes the final temperature is of the order of
 $\sim$10 mK, which is low enough for achieving Coulomb
 crystallization ($\Gamma\sim 250$~\cite{Schiffer:2002}). Imaging of
 the fluorescence from the trapped ions is achieved using an image
 intensified CCD camera placed above the trap.

 \begin{figure}[htbp]
   \centering
   \includegraphics[width=8cm,clip=true]{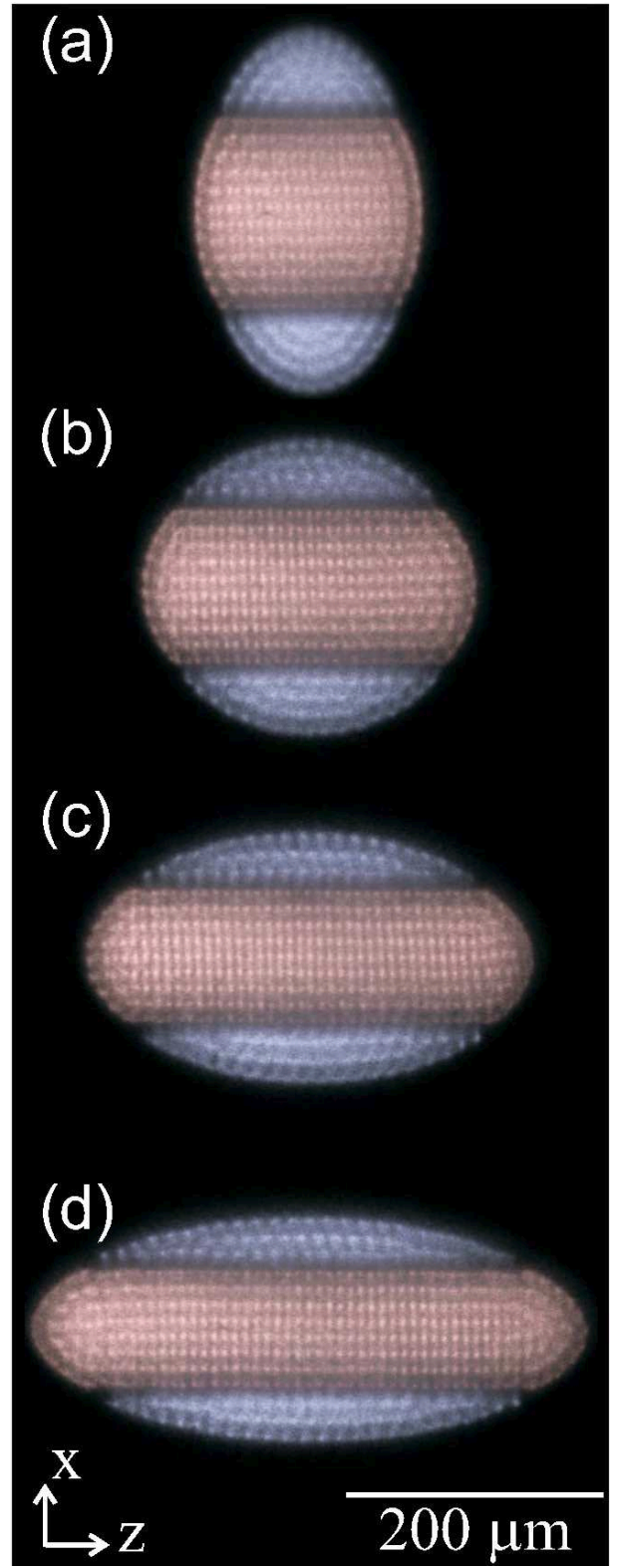}
   \caption{(color online) Images of two-species ion Coulomb crystals
     containing $\sim$1500 $^{40}$Ca$^+$ ions (red) and $\sim$2000
     $^{44}$Ca$^+$ ions (blue) at different settings of the end-cap
     potential, $U_\mathrm{end}$. The camera exposure time is $\sim$100 ms and the trap potentials are $U_\mathrm{rf} = 540$ V and
     (a) $U_\mathrm{end}= 46.1$ V, (b) $U_\mathrm{end}= 30.2$ V, (c)
     $U_\mathrm{end}= 20.0$ V, and (d) $U_\mathrm{end}= 13.8$ V,
     respectively.  }
   \label{fig:figure2}
 \end{figure}
 Images of two-species Coulomb crystals consisting of $^{40}$Ca$^+$
 and $^{44}$Ca$^+$ ions are presented in Fig.~2 for various settings
 of the static voltage $U_\mathrm{end}$ on the end-electrodes.  Since
 the ions are only fluorescing when they are directly laser cooled,
 the two isotopes can be imaged separately by alternatingly blocking
 (about 1 Hz rep. rate) the 850 nm ($^{40}$Ca$^+$ laser cooling off)
 and the 866 nm laser ($^{44}$Ca$^+$ laser cooling off) using
 mechanical shutters. The presented  combined images are subsequently
 created with a red color coding for $^{40}$Ca$^+$ and blue for
 $^{44}$Ca$^+$.  Due to sympathetic
 cooling~\cite{Larson:1986,Bowe:1999}, the crystal retains its shape
 and structure during the alternating laser cooling sequence.  As
 expected~\cite{Hornekaer:2001}, the lightest isotope $^{40}$Ca$^+$ is
 located as a cylindrical core closest to the trap axis, surrounded by
 the heavier $^{44}$Ca$^+$. It is immediately clear from the images in
 Fig.~2 that the projection of the actual three-dimensional structure
 of the $^{40}$Ca$^+$ ions is a two-dimensional rectangular lattice
 aligned with the trap axis. Since the depth of focus of the imaging
 system ($\sim$50 $\mu$m) is several times the inter-ion distance, we
 conclude the rectangular structure in the images must originate from
 a three-dimensional long-range ordering~\cite{Mortensen:2006}.  In
 contrast to our previous observations of long-range order in
 spherical one-component ion crystals~\cite{Mortensen:2006}, where the
 orientation of the observed metastable ($\sim$100 ms) structures
 seemed to be arbitrary, the rectangular structures presented here are
 very persistent ($\sim$10 s) and always oriented the same way.
 Accordingly, the presence
   of the surrounding $^{44}$Ca$^+$ ions apparently has significant influence on the formation and appearance of the observed long range
   structure in the $^{40}$Ca$^+$ part of the crystal.
 The image sequence in Fig.~2 illustrates additionally that
 despite changes in the outer shape of the $^{40}$Ca$^+$ core, the
 observed rectangular lattice of the ions is preserved, indicating
 that the observed long-range ordered structure is rather stable to
 changes in the boundary conditions of the crystal.

 \begin{figure}[htbp]
   \centering \includegraphics[width=8cm,clip=true]{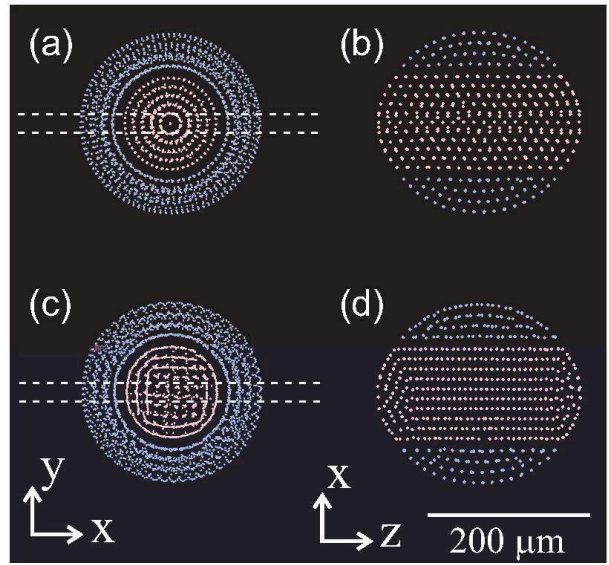}
   \caption{(color online) MD simulation data of a two-species ion Coulomb crystal containing 1500
     $^{40}$Ca$^+$ ions (red) and 2000 $^{44}$Ca$^+$ ions (blue) using
     a pseudo potential approximation ((a) and (b)) and using the full
     rf potential ((c) and (d)), respectively. While (a) and (c) show
     a projection of all the ion positions onto the $xy$-plane, (b)
     and (d) present the data points contained in a slice of thickness
     24 $\mu$m through the crystal center projected to the $xz$-plane.
     The slices are indicated by white dashed lines in (a) and (c).
     Trap potentials are $U_\mathrm{rf}= 540$ V and $U_\mathrm{end}=
     33$ V.  }
  \label{fig:figure3}
 \end{figure}

 In order to understand the observations, a series of molecular
 dynamics (MD) simulations of two-species crystals with the same
 number of the two calcium isotope ions as in the crystals shown in
 Fig.~2 have been performed. In Figs.~3(a) and 3(b), results from one
 simulation using a pseudo-potential corresponding to the trapping
 parameters of Fig.~2(b) is presented. As in the experiments, a clear
 radial separation of the two isotope ions is observed. Furthermore
 the radial projection (Fig.~3(a)) clearly reveals that the
 $^{40}$Ca$^+$ part of the crystal is organized in concentric
 cylindrical structures, resembling the structure of an infinitely
 long 1D cylindrically symmetric confined ion
 crystal~\cite{Hasse:1990,Hornekaer:2001}. In Fig.~3(b), a projection
 corresponding to the focal region of the imaging system is shown.
 Neither this nor other sections, e.g., in the $yz$-plane, lead to
 projection images with rectangular structures. However, when the full
 rf potential is used in the MD simulations, some regular ordering in
 the central component of the crystal does appear, as is evident from
 the results presented in Figs.~3(c) and 3(d) for a specific phase of
 the rf field. From Fig.~3(d), it is seen that indeed a rectangular
 projection image is expected when the rf-quadrupole field, which
 breaks the rotational symmetry, is included in the simulations. Even
 when averaging over all phases of the rf field, the rectangular
 structure persists. However, some blurring of the position of the
 ions along the $x$-axis, as is seen in the images of Fig.~2, is
 found.  Analysis of a much simpler two-ion system in a linear rf trap
 has previously shown similar preferred orientation effects with
 respect to the rf quadrupole field axes~\cite{Drewsen:2000}.

 \begin{figure}[htbp]
   \centering
   \includegraphics[width=8cm,clip=true]{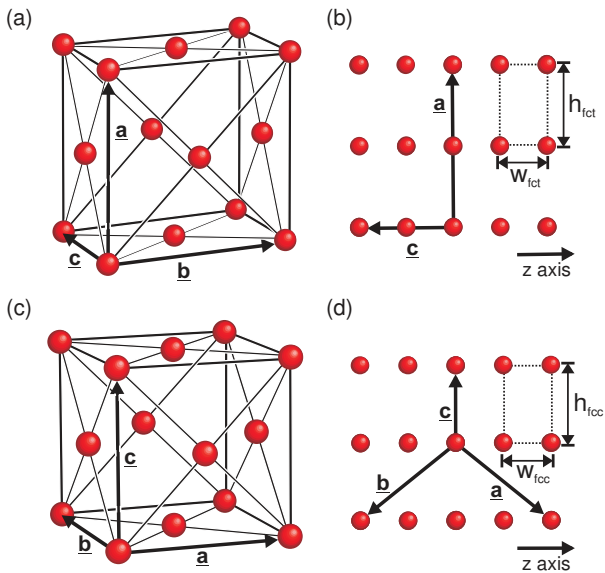}
\caption{Fct lattice cell (a) with projection along the $a$ vector (b). Fcc lattice
     cells (c) with projection along the [211]-direction (d).}   
   \label{fig:figure4}
 \end{figure}

 A closer analysis of the simulation results presented in Figs.~3(c)
 and 3(d) shows that the $^{40}$Ca$^+$ ions in the core organize
 themselves in a long-range ordered structure in the form of a
 face-centered tetragonal (fct) lattice as illustrated in Fig.~4(a),
 with the sidelengths related by $a=b= \sqrt{3}c$. The rectangular
 projection presented in Fig.~3(d) is obtained when the fct structure
 is viewed along the $b$ vector as illustrated in Fig.~4(b). This
 rectangular projection has a height to width ratio of
  $h_\mathrm{fct}/w_\mathrm{fct}= \sqrt{3} \simeq 1.73$, which is not exactly the
 same as the  $h/w =1.62 \pm 0.07$ observed in the experiments.
 In fact, the observed rectangular structure is more in agreement
 with a face-centered cubic (fcc) structure viewed along the [211]
 direction (see Figs.~4(c) and 4(d)), which would lead to
  $h_\mathrm{fcc}/w_\mathrm{fcc}= \sqrt{8/3} \simeq 1.63$. The reason for the
 deviation between the simulated results and the actually observed
 structures is probably that the difference in the potential energies
 of the two structures in the rf potential is very small, as is
 well-known to be the case for various long-range ordered structures
 in infinite systems without the presence of rf
 fields~\cite{Dubin:1989}.  In recent single component experiments
 both body-centered cubic (bcc) and fcc-like structures were indeed
 observed~\cite{Mortensen:2006}, but in contrast to the two-species
 results above, no fixed orientation with respect to the trap axis was
 found. Another point supporting that the observed structure is a fcc
 structure is the ion density. Assuming that the observed projection
 images of $^{40}$Ca$^+$ ions in Fig.~2 are actually fcc structures
 observed along the [211] direction, the ion density must be
 $n_\mathrm{fcc} = (3.8 \pm 0.4) \times 10^8$ cm$^{-3}$. In
 comparison, the $^{40}$Ca$^+$ ion density calculated from the trap
 parameters is $n_\mathrm{theo,40} = (4.3 \pm 0.3) \times 10^8$, in
 good agreement with the fcc assumption.

  Very stable aligned crystal structures as those discussed
   above, may find many future applications. For instance, for
   cavity QED studies, the situation where the axial periodicity of
   the Coulomb crystal is an  integer multiple of the node spacing of the
   standing wave cavity field is very interesting, since the
   effective coupling of the atomic ensemble to the light field can be
   controlled by shifting the position of the whole crystal.

  In conclusion, very persistent  three-dimensional long-range ordered structures with
   one specific orientation have been observed in two-species ion
   Coulomb crystals in a linear Paul trap. MD
   simulations strongly indicate that these characteristics are a
   consequence of the co-axial cylindrical symmetry of the central ion
   component of the Coulomb crystal and the radio frequency quadrupole
   trapping field.
 

 We acknowledge financial support from the Carlsberg Foundation.
 

\end{document}